\documentclass[twocolumn]{webofc}

\usepackage{booktabs}
\usepackage{hyperref}
\usepackage{arydshln}
\usepackage{fvextra}
\usepackage{tikz}
\usepackage{enumitem}
\usetikzlibrary{positioning, arrows.meta, fit, calc}
\usepackage{xcolor}
\definecolor{docgreen}{HTML}{2E7D5A}
\definecolor{docfill}{HTML}{E6F2EA}
\definecolor{liveorange}{HTML}{C75B25}
\definecolor{livefill}{HTML}{FBE6D8}
\definecolor{agentblue}{HTML}{0072B2}
\definecolor{agentfill}{HTML}{DCEAF8}

\makeatletter
\renewcommand\@FirstNameAuthorFont{\normalfont}
\renewcommand\@LastNameAuthorFont{\normalfont}
\makeatother

\title{Archi: Agentic Operations at the CMS Experiment}
\titlerunning{Archi}
\authorrunning{Archi authors}
\author{%
  \firstname{Pietro} \lastname{Lugato}\inst{1,2} \and
  \firstname{Luca} \lastname{Lavezzo}\inst{1,2} \and
  \firstname{Jason} \lastname{Mohoney}\inst{1} \and
  \firstname{Hasan} \lastname{Ozturk}\inst{2} \and
  \firstname{Muhammad Hassan} \lastname{Ahmed}\inst{2} \and
  \firstname{Juan Pablo} \lastname{Salas}\inst{2,3} \and
  \firstname{Viphava} \lastname{Ohm}\inst{2,3} \and
  \firstname{Krittin} \lastname{Phornsiricharoenphant}\inst{2} \and
  \firstname{Gabriele} \lastname{Benelli}\inst{2,4,5} \and
  \firstname{Mariarosaria} \lastname{D'Alfonso}\inst{1,2} \and
  \firstname{Manasvita} \lastname{Joshi}\inst{6} \and
  \firstname{Warren} \lastname{Nam}\inst{1} \and
  \firstname{Aron} \lastname{Soha}\inst{2,4} \and
  \firstname{Samantha} \lastname{Sunnarborg}\inst{2,5} \and
  \firstname{Austin} \lastname{Swinney}\inst{6} \and
  \firstname{Jack} \lastname{Tucker}\inst{1} \and
  \firstname{Dmytro} \lastname{Kovalskyi}\inst{1,2} \and
  \firstname{Tim} \lastname{Kraska}\inst{1} \and
  \firstname{Christoph} \lastname{Paus}\inst{1,2}
}
\institute{%
  Massachusetts Institute of Technology, Cambridge, MA, USA \and
  CMS Collaboration, CERN, Geneva, Switzerland \and
  University of Wisconsin-Madison, Madison, WI, USA \and
  Fermi National Accelerator Laboratory, Batavia, IL, USA \and
  Brown University, Providence, RI, USA \and
  Harvard University, Cambridge, MA, USA
}

\begin{document}

\abstract{%

We present Archi, an open-source, end-to-end framework for scientific collaborations that combines the systematic ingestion and organization of heterogeneous data sources with the deployment of configurable, private, and extensible agents that retrieve and reason over them.
An instance of Archi has been deployed for the Computing Operations team of the CMS experiment at CERN's LHC since February 2026 as a support agent for technical operators, offering retrieval and analysis capabilities by combining documentation, historical data, and live monitoring systems.
We evaluate the system on operator feedback and a question set collected from production usage, graded by human and automated panels.
The system proves effective at operational tasks, resolving real-world queries posed by CMS operators.
We also observe that locally-hosted, open-weight models perform competitively, enabling fully private management of sensitive data.

}

\maketitle
\section{Introduction}
\label{sec:introduction}

Scientific collaborations, such as the Compact Muon Solenoid (CMS) experiment at CERN's Large Hadron Collider (LHC), have developed increasingly complex webs of knowledge infrastructure over the years.
This has resulted in significant problems for scientists, engineers, and operators working for these experiments.
Retrieving relevant and accurate information over heterogeneous, scattered, and inevitably conflicting or outdated sources is difficult and time-consuming, often forming a bottleneck in the running of experiments and progress of scientific work.

Recent advances in the fields of natural-language processing and artificial intelligence have produced techniques for the systematic ingestion, organization, and retrieval of information across sources, offering a promising avenue for addressing these challenges.
Building on these developments, we present Archi\footnote{Code: \url{https://github.com/archi-physics/archi}; project site: \url{https://archi-physics.github.io/website/}.}, a framework for data systematization and agentic operations in scientific collaborations.
Archi combines a unified ingestion pipeline for heterogeneous data sources, a retrieval-augmented generation (RAG) layer over the resulting knowledge base, and a deployment system for customizable, task-specific agents.

Beyond these components, scientific collaborations impose requirements that generic retrieval and agentic systems do not address, and Archi is designed around them.
For data privacy, it offers flexible support for running in-house on open-weight models, together with anonymization of sensitive information.
For reliability, it grounds answers in the collaboration's own data rather than in the model's memory, and exposes the steps it takes.
It is also accessible, so groups can deploy it despite lacking relevant personnel or expertise, and modular, so it can accommodate the data and requirements unique to different experiments.
Finally, Archi is open-source and collaborative, so features developed for one deployment can benefit all, as does the core effort to maintain it and keep up to date with the latest technologies.

The rest of the paper is structured as follows.
The Archi framework is introduced in Section~\ref{sec:archi-framework}.
We then turn to a concrete example of the aforementioned problems in the context of Computing Operations at the CMS experiment, Section~\ref{sec:compops}, and describe the Archi deployment for this use-case, Section~\ref{sec:compops-agent}.
Using data collected from this deployment, we evaluate the agent's performance both via LLM judges and a panel of CMS domain experts, assessing the quality of the system to solve real-world operational queries across a suite of metrics, and studying how different underlying LLMs, and the inclusion of different tools and data sources affect the results, Section~\ref{sec:evaluation}.
Related work, deployments of the Archi framework in other CMS groups, and future directions are outlined, Section~\ref{sec:discussion}, before concluding, Section~\ref{sec:conclusions}. 

\section{The Archi Framework}\label{sec:archi-framework}

\begin{figure*}[!htbp]
\centering
\includegraphics[width=0.95\textwidth]{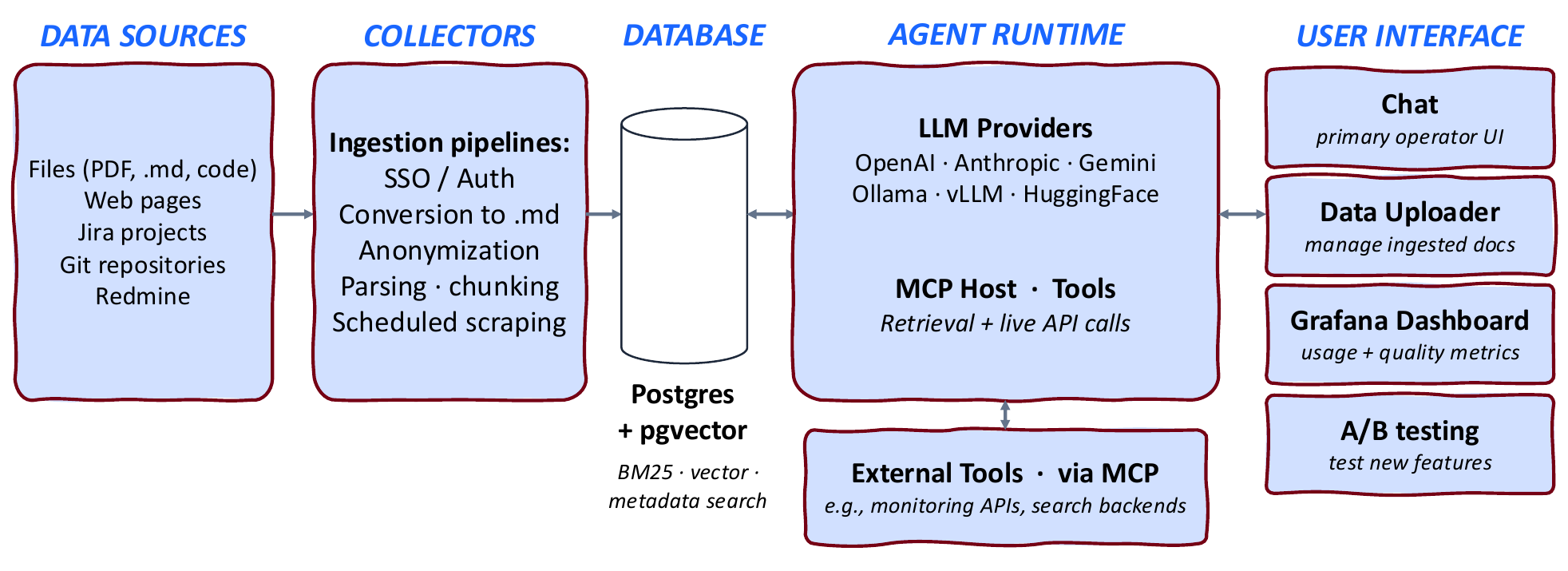}
\caption{Archi's architecture: data sources feed ingestion collectors that write to a PostgreSQL + \texttt{pgvector} store, which an agent runtime queries via BM25, vector, and metadata search and from which it composes LLM and tool calls, fronted by an operator-facing user interface with a chat, data viewer and uploader, monitoring, A/B testing, and more. Sources and tools are added according to the use case; those shown in the diagram are more relevant to CMS Computing Operations.}
\label{fig:framework}
\end{figure*}

Archi is an open-source, end-to-end, Python framework for collecting and organizing heterogenous data sources, and deploying custom agents that can retrieve and reason over them. 
Figure~\ref{fig:framework} shows an overview of the architecture.
The Archi documentation\footnote{https://archi-physics.github.io/archi/} elaborates and expands upon what is outlined in this Section.

\subsection{Data manager: data sources, collectors, database}

The data manager comprises all steps taken to collect, process, and store data in a manner that is optimized for agent retrieval. \textit{Collectors} retrieve data from specific \textit{data sources}. Archi supports collectors for generic web scraping, JIRA tickets, git repositories, and local files, among others. Collectors are modular, one can add a collector for a new data source as needed. Gathered data gets passed to a common ingestion pipeline, which performs any of the following, as configured: conversion to markdown format for better legibility by LLMs, anonymization for protecting personal information, and chunking. These documents are then stored both as text and as vectors using an embedding model of choice into a postgres \textit{database} via the \texttt{pgvector} extension. Many aspects of this pipeline are configurable to the user, including but not limited to setting of scheduled updates per source, choice of search algorithms, and chunk properties.

\subsection{Agent runtime}
A common environment is provided for \textit{agent runtime} that supports all major LLM providers, as well as MCP~\cite{10.1145/3796519} servers and custom tools.
Each deployment can define its own agent class with the LangGraph framework \cite{langgraph} combined with custom wrappers.
The agent classes can inherit from a base ReAct \cite{yao2023reactsynergizingreasoningacting} agent, single-shot RAG, or be entirely custom.
Choice of LLM, prompts, and tools can be passed to this class, but need not be decided here.

\subsection{User interface}
The system is designed such that any \textit{service}, applications using the data manager and agents as a backend, can easily be deployed via the command line interface (CLI).
By default, Archi provides a chat interface with authentication requirements decided by the admin.
Via the chat interface, users can define their own agent specs within a range defined by the admin: adjusting the prompt, enabling tools, and selecting LLM provider and model.
There is also a separate page for admins to view and manage data sources.
Users can provide feedback directly through the chat interface.
A Grafana dashboard is provided to admins to monitor usage and quality metrics, while a separate status board with alerts for downtime or new versions is accessible to all users.
A/B testing is available to test new features, different models, etc.
\textit{Role-based access control (RBAC)} is implemented to make for easy assignment of privileges to different sets of users.

\subsection{Deploying Archi}
The system is designed to be easy to set up and launch.
Archi provides a CLI which can be used to launch the full, containerized deployment provided a few required options are passed.
At least one \textit{agent spec}, a markdown file with exposed tools and system prompt, must be passed to serve as the default.
A YAML \textit{configuration file} defines the options for the data manager and services.
Secrets for accessing data sources or services are also passed through the CLI. 
The system can run on Podman or Docker with any network configuration, on CPU or GPU, and supports various options for development and testing.

\subsection{Benchmarking and evaluation}
Archi includes an in-house evaluation framework to ground improvements in trackable metrics.
Provided via the CLI with a reference set of questions, answers, and a list of relevant sources per corresponding question, the framework supports two evaluation modes: measuring retrieval accuracy or recall against the relevant sources, and grading the generated answers with an LLM judge.

\subsection{Open source and privacy}
The Archi framework is open source, and provides the means for deployments to run locally and enforce privacy measures.
The anonymization pipeline helps strip ingested data of personal information.
It uses named-entity recognition (via spaCy~\cite{spacy}) to remove personal names, together with configurable regex patterns for emails, usernames, and email-style greetings and sign-offs.
Flexible support for open-weight models including Ollama, vLLM, and HuggingFace, plus the ability to run on GPUs on the same machine where Archi runs, or to ping another local model, ensures that the deployment of Archi is agnostic to local model setups.
During deployment, authentication and RBAC make sure that control over the ingested data is accessible only to approved persons.

\section{CMS Computing Operations}
\label{sec:compops}

The most mature deployment of the Archi framework is for the CMS Computing Operations (CompOps)~\cite{cms_compops}.
The group is responsible for operating the CMS offline computing chain: transforming the raw output of the CMS detector into usable scientific data, producing Monte Carlo simulation datasets, and making these data available to physicists worldwide.
Within CompOps, three sub-teams share this work: the Tier-0 team converts recorded events from the CMS detector into the long-lived raw data files that all subsequent analysis depends on, archives them to tape, and runs the first reconstruction passes, a fast pass on a subset of events for early calibration, and data-quality feedback, followed by a full pass over the complete dataset using the latest calibrations.
The Production \& Reprocessing team drives Monte Carlo simulation and data reprocessing campaigns through the CMS Workflow Management system~\cite{cms_wmcore}.
Data Management~\cite{cms_dm2025} oversees more than an exabyte of data distributed across more than sixty storage sites on the Worldwide LHC Computing Grid (WLCG)~\cite{wlcg}, executing central data transfers, managing placement and deletion policies, and operating the AAA (Any Data, Anytime, Anywhere) service~\cite{bloom2015datatimeanywhereglobal} for remote data access. All of this is operated by a core CompOps team of roughly fifteen people, working alongside several sister teams.

The tools and knowledge that CompOps operators rely on are scattered across many systems.
At the core lies a stack of production services: Rucio~\cite{rucio} orchestrates data management, FTS~\cite{fts} executes the underlying file transfers, HTCondor~\cite{htcondor} pools run the batch workload, and the CMS Workflow Management system (WMCore)~\cite{cms_wmcore} drives production campaigns.
Operators interact with each through its own APIs, clients, web interfaces, and service-specific monitoring dashboards, inspecting directly the underlying deployments and databases (e.g., Kubernetes, Oracle) when deeper diagnostics are required.
Operators work across CERN, Fermilab, and other collaboration centers in different time zones.
Issues that exceed operator-level expertise are escalated to domain experts through ticketing systems (JIRA, GGUS, ServiceNow).
Operational metrics and events are stored in OpenSearch and visualized through Grafana~\cite{cms_monit}, while low-level logs remain fragmented across local and shared file systems, as well as service-specific logging backends.
Documented procedures are hosted on CMS wiki pages (TWiki, Gitlab).

Day-to-day, CompOps operators and engineers must navigate this scattered landscape to perform their work: monitoring live operational state, debugging anomalies, answering tickets, and contacting domain experts.
Even a single task typically spans five or more sources. Diagnosing a transfer anomaly, for example, requires cross-referencing Rucio's monitoring dashboards, its API, its logs, FTS logs, and the affected site's GGUS tickets.
The cumulative cost is significant.
Further, hard-won operational knowledge leaves with each rotation, and new operators typically require months to onboard.

\section{Archi Deployment for CMS Computing Operations}
\label{sec:compops-agent}

The challenges outlined in the previous Section, emblematic of many of those faced by modern scientific collaborations, make CMS Computing Operations a natural candidate for Archi, which can systematize its data and provide agents to intelligently retrieve and act on it.
Here we describe the Archi deployment for this team: its agent, current data sources and tools it draws from, the production setup, and representative workflows drawn from live use.

\subsection{The Agent}

The Archi deployment for CMS CompOps runs an agent built on the ReAct pattern~\cite{yao2023reactsynergizingreasoningacting}.
The agent alternates a reasoning step (which emits a textual trace) with a tool-call step; when the reasoning step decides no further tools are needed, it emits the final answer with the accumulated tool-call trace.
The system prompt steers the agent toward exploratory tool calls, aggregation queries when counting, and fetch-by-ID fallback when a search returns only summaries.

\subsection{Data Sources and Tools}

Archi exposes two classes of data sources to the agent: a periodically ingested document corpus through Archi's own data-manager, and external services queried live on demand via custom MCP tools.

The ingested corpus comprises approximately ten thousand documents.
Web pages containing documentation, procedures, and other useful information are scraped to plain text files; most of these pages are private, CERN Single Sign-On (SSO) protected, and a custom Selenium-based scraper is employed.
Git repositories of documentation (e.g. MkDocs) are also ingested.
Operational knowledge and correspondences are brought in from the JIRA ticketing system with issues and comments fetched periodically from several configured projects.
The agent is provided with tools to search through this ingested corpus, which include semantic search, BM25 keyword search, grep-like content search, and metadata filtering.

The on-demand tools query external services for live operational state: the Rucio events monitoring index, Rucio client, the HTCondor jobs monitoring index, and DeepWiki~\cite{cognition_deepwiki_2025} as an auxiliary code-navigation aid for the CompOps codebases.
Every tool call is recorded in the trace surfaced to the operator.

\subsection{Production Deployment}
\label{sec:deployment}

Archi has run as the CMS CompOps operator chat assistant since February 2026 on CMS infrastructure at CERN, serving the operations team.
The agent is accessible via a web interface, hosted on the CERN network, and protected by SSO authentication.
The chat interface maintains turn-by-turn conversation history, lists source documents per response, and exposes the agent's tool-call trace.

\paragraph{Privacy and guardrails.}
The deployed pipeline runs on OpenAI's GPT-5 family, currently GPT-5.5, as well as open-weight models.
Use of OpenAI models in the production deployment follows the CERN AI data privacy guidelines \footnote{\url{https://privacy.web.cern.ch/artificial-intelligence-ai}}, and users are pointed to a dedicated CompOps privacy notice; open-weight models run on CMS-internal infrastructure with no external API calls.

JIRA tickets are anonymized at ingest by stripping personal identifiers.
Access control is role-based; the tool set is configurable per use case; and ingestion is rate-limited to protect the upstream services it scrapes.
Authentication to the live tools uses a service account with read-only permissions, and every user with access to the agent already holds read-only access to the underlying data sources.
The agent's capabilities are therefore a subset of the user's own.

\paragraph{First impressions from production.}
Between 13~February and 29~April 2026, the deployed assistant served 20 distinct users (of 22 with access) across 393 conversations and 598 user messages, with a mean response time of 107~seconds and 99 explicit voluntary feedback events (70\% positive, 30\% negative).
The dominant negative-feedback category was \emph{missing capabilities}; specifically, requests for live data sources the agent does not yet expose, ahead of retrieval bugs, anonymization frustrations, and tool misuse.

\paragraph{Examples from production.}
 
Four workflows distilled from the recorded operator traffic illustrate typical use cases: three successful, one exposing a shortcoming.

\emph{Success: transfer failure diagnosis.} An operator pastes the Rucio Rule ID of a failing data transfer and asks Archi for the cause and a solution. The agent queries the Rucio client to enumerate the failing files and their destination sites, then aggregates the corresponding records in the Rucio events monitoring index to identify the dominant failure modes. Conditioned on the top error, it searches JIRA for similar past occurrences and retrieves the matching section of the operations documentation. It returns the leading error with its likely causes, the most recent matching ticket, and the runbook section for the failure mode.

\emph{Success: efficiency drop investigation.} An operator reports a global drop in CPU efficiency and asks which workflows are responsible. The agent aggregates HTCondor job monitoring records by workflow, restricted to completed jobs with low \texttt{CpuEff}, and inspects each top contributor's CPU request, task mix, and input locality; it returns a ranked list with per-workflow diagnostics (over-provisioned cores, non-CPU-bound service steps, offsite reads) alongside suggested mitigations based on retrieved tickets and documentation.

\emph{Success: procedure lookup.} A Tier-0 operator asks how to reprocess five runs whose streamer files are still on disk. The agent retrieves the Tier-0 replay procedure from the TWiki and matching JIRA tickets, and queries DeepWiki against the T0 codebase to verify the replay-deployment scripts; returns the step-by-step procedure and the key differences from production.

\emph{Failure: environment mismatch.} An operator asks for a Rucio CLI command to list a dataset's file paths. The agent proposes one from the upstream Rucio documentation, but the installed Rucio CMS client rejects an argument added only in a newer release; when the operator pastes the error back, the agent gives a working alternative. The root cause is that the agent has no visibility into the Rucio CLI version in the CMS environment, so it cannot disambiguate documentation pages describing different releases.

\section{Evaluation}
\label{sec:evaluation}

In addition to collecting the operators' feedback, we evaluate the system's performance directly on real user queries collected from the deployed production traffic.
The agent is run offline using different configurations on the user queries.
Evaluation focuses on four questions: (Q1) whether Archi can address real-life operational questions, (Q2) how much CMS project context is needed to achieve high rubric-scored answer quality, (Q3) whether iterative tool use improves over a single retrieval pass, and (Q4) how open-weight and frontier models compare.
Evaluation is performed by separate panels of CMS domain experts and automated judges.

\subsection{Setup}
\label{sec:eval_setup}

\begin{figure}[t]
\centering
\includegraphics[width=\linewidth]{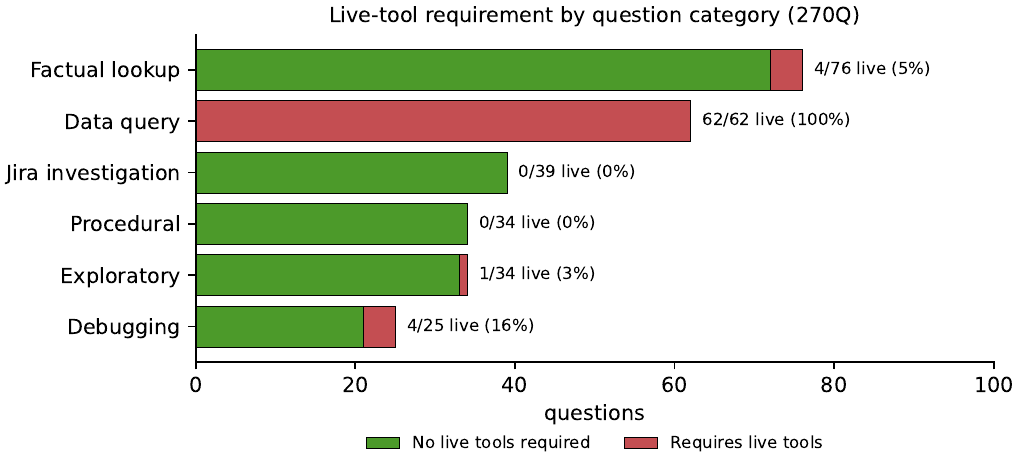}
\caption{
Live-tool requirement per question category in the 270Q set.
The audit separates questions answerable from static project context from questions requiring current operational state.
}
\label{fig:eval_workload_live_tools_270q}
\end{figure}

We report two question sets collected from production traces and curated by an experienced CMS operator. 
A set of 63 questions (63Q) set is used for both human and automated grading.
An expanded set of 270 questions (270Q), a superset of the 63Q, is used only for automated grading.
As shown in Figure~\ref{fig:eval_workload_live_tools_270q}, the questions represent a wide range of operational issues: factual lookup, live data queries, ticket investigations, procedure retrieval, exploratory investigation, and debugging.
Of the 270 questions, 199 require no live tools, whereas 71 require current live operational state.

Four Archi pipeline configurations are compared, together with the deployed production system as a reference. \textit{Bare} sends the
question directly to the LLM. \textit{Single-shot RAG} retrieves once from the indexed CMS corpus and answers from the retrieved context.
\textit{Agent-no-live} is an iterative agent restricted to the static corpus.
\textit{Agent-with-live} is the same iterative agent with live operational tools; this is the configuration used in the current CMS deployment.
The static corpus and live-data tools are the same as those used in the real deployment described in Section~\ref{sec:compops-agent}.

We compare open-weight and frontier models as backends to the Archi agents by comparing the latest Qwen and OpenAI models.
Qwen3.6-27b and Qwen3.6-35b-a3b runs used vLLM on two NVIDIA H200 GPUs, 16 CPUs, and 400~GB RAM, serving the models at native \texttt{bf16} precision with no quantization.
Archi services ran on CPU nodes with eight CPUs and 32~GB RAM.
GPT-5.5 generation used the OpenAI API.
The human panel was comprised of CMS domain experts and operators.
The panel scored blinded responses for correctness and usefulness on 1--5 scales and ranked responses per question, with ties allowed. Automated LLM judges scored a wider range of configurations than the human panel could cover.
On the 63Q set we average four LLM judges as a cross-check against single-judge bias; on the larger 270Q set we use the source-free GLM-5.1 judge alone.
Every judge scores each configuration from the generated answer alone, with no reference answer, on a reference-free rubric of relevance, completeness, specificity, and helpfulness, and we report the unweighted mean of these dimensions.

\subsection{Results}
\label{sec:eval_63q}

Results for the human and automated grading on the 63Q set are shown in Tables~\ref{tab:human_63q} and~\ref{tab:llm_judge_63q_core4}.

\begin{table}[!ht]
\centering
\caption{Human ratings on 63Q. Correctness and usefulness are 1--5 means;
\emph{Avg.}\ is their unweighted mean, with its standard error.
Rank is lower-is-better. $n_q$ is the number of questions scored and $n_r$ the number of individual ratings collected.}
\label{tab:human_63q}
\scriptsize
\setlength{\tabcolsep}{2pt}
\begin{tabular}{@{}p{2.25cm}rrrrrr@{}}
\toprule
Configuration & Correct. & Useful. & Avg. & Rank & $n_q$ & $n_r$ \\
\midrule
GPT-5.5 Agent-with-live & 4.25 & 4.07 & \textbf{4.16\,$\pm$\,0.09} & \textbf{1.95} & 63 & 265 \\
Qwen3.6-35B Agent-with-live & 3.87 & 3.72 & 3.80\,$\pm$\,0.12 & 2.31 & 63 & 265 \\
Qwen3.6-27B Agent-with-live & 3.87 & 3.72 & 3.79\,$\pm$\,0.12 & 2.25 & 63 & 265 \\
Qwen3.6-27B Agent-no-live & 3.83 & 3.65 & 3.74\,$\pm$\,0.14 & 2.28 & 63 & 265 \\
Qwen3.6-27B Single-shot RAG & 2.58 & 2.33 & 2.45\,$\pm$\,0.17 & 3.62 & 53 & 244 \\
\bottomrule
\end{tabular}
\end{table}

\begin{table}[!ht]
\centering
\caption{Four-judge automated ratings on 63Q. Mean is the unweighted mean over
relevance, completeness, specificity, and helpfulness, with its standard error.}
\label{tab:llm_judge_63q_core4}
\scriptsize
\setlength{\tabcolsep}{3.5pt}
\begin{tabular}{@{}p{2.25cm}rr|rrrr@{}}
\toprule
Configuration & Mean & $n_q$ & Rel. & Comp. & Spec. & Help. \\
\midrule
GPT-5.5 Agent-with-live & \textbf{4.59\,$\pm$\,0.18} & 63 & \textbf{4.99} & \textbf{4.77} & \textbf{4.14} & \textbf{4.47} \\
GPT-5.3 Agent-with-live & 4.32\,$\pm$\,0.15 & 53 & 4.67 & 4.28 & 4.12 & 4.23 \\
Qwen3.6-27B Agent-with-live & 4.32\,$\pm$\,0.23 & 63 & 4.82 & 4.50 & 3.82 & 4.14 \\
Qwen3.6-35B Agent-with-live & 4.13\,$\pm$\,0.23 & 63 & 4.69 & 4.40 & 3.55 & 3.88 \\
Qwen3.6-27B Agent-no-live & 4.10\,$\pm$\,0.24 & 63 & 4.58 & 4.24 & 3.61 & 3.96 \\
Qwen3.6-27B Single-shot RAG & 3.44\,$\pm$\,0.22 & 63 & 4.09 & 3.30 & 3.15 & 3.23 \\
\bottomrule
\end{tabular}
\end{table} 

The LLM and the human panels are aligned, showing similar ordering for overlapping configurations.
Both rank the GPT-5.5 agent first ahead of all open-weight models, keep context-enabled agents ahead of the RAG control, and prefer agents with live-tools enabled.
The configuration GPT-5.3 Agent-with-live was the production version during the window during which feedback was collected.

The dimension columns of the automated evaluation show that relevance is high for most agent configurations.
The separation comes from completeness, specificity, and helpfulness: stronger systems give operators more complete and more actionable guidance without adding unsupported detail.

The broader automated matrix measures pipeline behavior on the 270Q set.
Table~\ref{tab:270q_common_success_core4} reports the common-success subset: 213 questions remain after excluding any question where a displayed configuration failed, produced no answer, or lacked a complete judge score.
This isolates answer quality from reliability.
The overlapping configurations between the two evaluation sets are ranked similarly, demonstrating alignment.

\begin{table*}[t]
\centering
\caption{
GLM-5.1 scores on the 213-question common-success subset.
Every listed configuration produced a non-empty answer and complete judge-dimension scores for these questions.
}
\label{tab:270q_common_success_core4}
\scriptsize
\setlength{\tabcolsep}{3pt}
\begin{tabular}{@{}lr|rrrr@{}}
\toprule
Configuration & Mean & Relevance & Completeness & Specificity & Helpfulness \\
\midrule
GPT-5.5 Agent-with-live & 4.63 & 4.94 & 4.86 & 4.21 & 4.51 \\
GPT-5.5 Agent-no-live & 4.56 & 4.91 & 4.62 & 4.26 & 4.46 \\
GPT-5.3 Agent-with-live & 4.39 & 4.71 & 4.44 & 4.12 & 4.30 \\
Qwen3.6-27B Agent-with-live & 4.38 & 4.88 & 4.72 & 3.77 & 4.13 \\
Qwen3.6-27B Agent-no-live & 4.33 & 4.83 & 4.59 & 3.77 & 4.13 \\
Qwen3.6-35B-A3B Agent-no-live & 4.28 & 4.79 & 4.50 & 3.82 & 4.00 \\
Qwen3.6-35B-A3B Agent-with-live & 4.27 & 4.83 & 4.62 & 3.61 & 4.01 \\
GPT-5.5 Single-shot RAG & 4.25 & 4.83 & 4.12 & 4.03 & 4.03 \\
Qwen3.6-35B-A3B Single-shot RAG & 4.09 & 4.72 & 3.89 & 3.93 & 3.80 \\
Qwen3.6-27B Single-shot RAG & 3.91 & 4.63 & 3.69 & 3.69 & 3.64 \\
Qwen3.6-27B Bare & 3.00 & 3.60 & 2.87 & 2.78 & 2.77 \\
Qwen3.6-35B-A3B Bare & 2.90 & 3.56 & 2.80 & 2.64 & 2.62 \\
\bottomrule
\end{tabular}
\end{table*}

\subsection{Findings}

\paragraph{Q1 -- Usefulness of the system.} Both the human panel and the automated judges found that Archi, in the configuration used for the real deployment, was able to provide correct (4.25/5) and useful (4.07/5) answers to real operator questions. This does not demonstrate that Archi can answer all issues an operator faces, only that it can answer those questions that they've asked it so far, which is a biased set, as for example questions asked by operators may have been informed by what they expected the system to know. 

\paragraph{Q2 -- Benefits of retrieved context.} CMS-specific project context is necessary: across all evaluations, the single-shot RAG configurations improved over the bare LLMs, and increased tool use and data access, first via agentic loops and then with the addition of live data sources, increased performance under all metrics.
Table~\ref{tab:qwen27_live_no_live_63q} shows the effect of using live tools on judged answer quality. \textit{Agent-with-live} is slightly ahead of\textit{ Agent-no-live }under the four-judge automated panel. The gap is larger among questions that require live tools and modest on those that do not, where unneeded live calls can dilute the context and slightly degrade answer quality.

\begin{figure*}[t]
\centering
\includegraphics[width=0.52\textwidth]{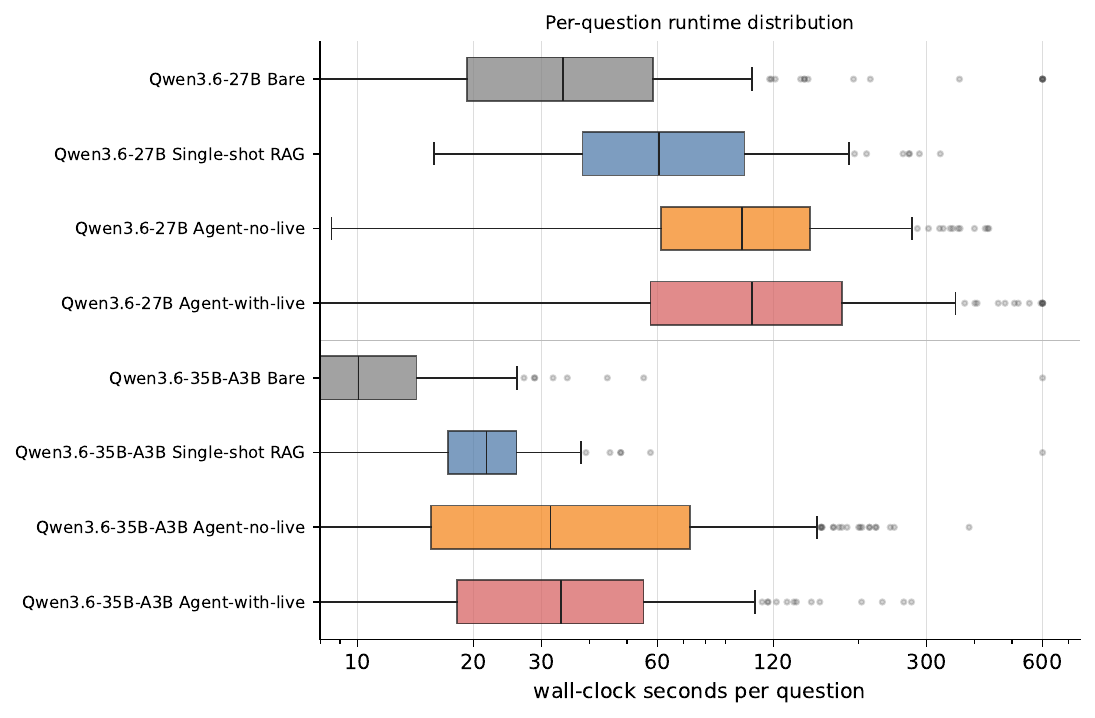}
\hfill
\includegraphics[width=0.40\textwidth]{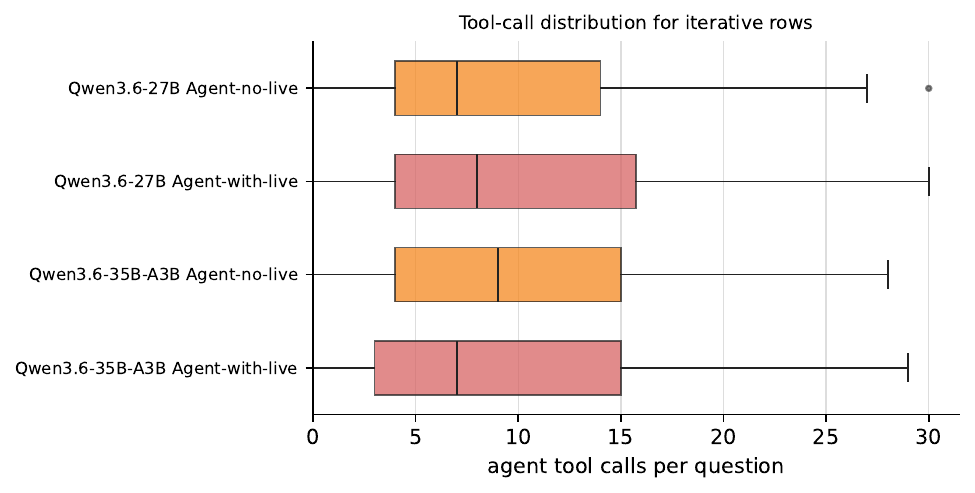}
\caption{Runtime and tool-call distributions for 270Q Qwen runs.
Runtime is wall-clock time per question on the local ORCD vLLM
setup; tool calls are benchmark trace events from iterative rows.}
\label{fig:270q_runtime_tools}
\end{figure*}

\paragraph{Q3 -- Agent versus single-shot RAG.} Agentic retrieval outperforms single-shot RAG in quality of answer.
However, the agent loop raises latency due to the multiple retrieval calls and reasoning steps. Figure~\ref{fig:270q_runtime_tools} reports the distribution of question
completion time. The figure is restricted to Qwen/vLLM rows; GPT API
latency is omitted because remote scheduling and network time are not
directly comparable to local H200 inference. Tool-call counts are shown
only for iterative rows, where they are recorded from benchmark trace
events.

\paragraph{Q4 -- Frontier versus open-weight models.} The best Qwen row remains below GPT-5.5 by 0.25 points on the common-success table and by 0.36 points in the human table.
It, however, matches GPT-5.3, released only about 4 months ago, suggesting that open-weight models are keeping pace with frontier models with a modest delay.
This best Qwen row is judged by the human panel to be on average correct (3.87) and useful (3.72) in its answers.
Considering the benefits of cost and privacy, it is a relevant finding for deciding which strategy to adopt.

\begin{table}[t]
\centering
\caption{Qwen3.6-27B Agent-with-live versus Agent-no-live on 63Q. Scores are four-judge
means over the four rubric dimensions.}
\label{tab:qwen27_live_no_live_63q}
\scriptsize
\setlength{\tabcolsep}{3pt}
\begin{tabular}{lrrr}
\toprule
Pipeline & All & Non-live Qs & Live Qs \\
\midrule
Agent-with-live & 4.32 & 4.39 & 4.15 \\
Agent-no-live & 4.10 & 4.24 & 3.73 \\
\bottomrule
\end{tabular}
\end{table}

\section{Discussion}
\label{sec:discussion}

Here we discuss related work to Archi, the additional deployments underway, and future work on incorporating relational knowledge and making Archi an active agent.

\subsection{Related Work}
\label{sec:discussion_related}

Similar agents have been developed for retrieval applications in large scientific collaborations.
AccGPT~\cite{accgpt2025} and chATLAS~\cite{chatlas2025} are recent HEP examples, and retrieval-augmented generation is now a standard approach for knowledge-intensive question answering~\cite{lewis2020rag,gao2023ragsurvey}.
Other operational assistants address similar workflows: GAIA~\cite{mayet2024gaia} connects an open-weight ReAct assistant to accelerator control-system tools at DESY, and the multi-lab logbook study~\cite{sulc2024logbooks} extracts operational structure from electronic logbooks across national labs.
Archi seeks to generalize this pattern across collaborations, providing a flexible, extensible, low-barrier-to-entry, and generic framework for such applications, and demonstrates its contribution in a deployment for CMS computing through its production-trace evaluation.

\subsection{Other Deployments}
\label{sec:discussion_deployments}

Two other CMS deployments use Archi outside Computing Operations.
WiSDQM, the DQM Shift Assistant, supports CMS shifters at the LHC
Physics Center at Fermilab. Its sources are data-quality monitoring
documentation, shift logs, daily reports, and live detector
information; its first users include Tracker DQM, Online DQM, and
Trigger shifters. The deployment keeps the same Archi structure but
replaces CompOps workflows with detector-monitoring questions during
data-taking. It tests Archi in shift work, where a useful answer is
often tied to detector state at that moment.

The CRAB~\cite{mascheroni_crab3_2015} User Analysis Assistant targets the CMS analysis-job
submission system. It ingests CRAB documentation and exposes a CRAB MCP
server as an agent tool, so questions about job status and failure logs
can be answered by live CRAB queries. This deployment is in
pre-production and is being benchmarked and stress tested. Both cases
reuse Archi by changing source lists, pipelines, and tool catalogs,
not by copying the CompOps corpus.

\subsection{Future Work}
\label{sec:discussion_future}

Passage retrieval is a poor fit for some operational relations. The
current retriever works well for procedure lookup, but it treats
tickets, sites, workflows, datasets, and transfer links as text
mentions instead of typed objects. The next Archi release adds a
knowledge-graph-backed layer for those objects, separating durable
metadata from time-dependent state and narrowing the search space for
multi-hop operational questions. The same layer will improve
provenance: when an answer combines a runbook, a ticket, and several
live queries, each substantive claim should point to the passage,
ticket field, or query result that supports it.

Archi is also moving from read-only assistance toward an active agent.
The current CompOps agent can inspect documentation, tickets, and operational state, but it cannot submit a workflow action, change a site setting, or update a ticket.
The active agent work requires constrained action schemas, dry-run previews, explicit operator approval, policy checks, and immutable audit logs.
These controls define the path from decision support to automating operations.
\section{Conclusion}
\label{sec:conclusions}

The Archi framework has been introduced as a system for organizing heterogeneous data sources and deploying agents that can retrieve over them to address operational problems common in large scientific collaborations.
Archi has run as the CMS Computing Operations operator chat
assistant since February 2026. It was deployed against a complex, heterogeneous-tool,
live-state operations workload and enabled retrieval over CMS documentation, code, seven Jira projects, and live-data calls all wired through the same pipeline abstraction.

On an evaluation set of questions sampled from six weeks of production operator traffic and graded by a panel of CMS experts and a set of automated judges, we demonstrate the effectiveness of the system to address operators' questions.
We find that added context, iterative tool use, and live-data each improve the correctness and usefulness of the answers, albeit at some cost of latency. 
We also find that while closed-weight, frontier models are shown to modestly outperform open-weight ones, the best open-weight model already matches a frontier model from a few months prior, a short lag whose cost and privacy benefits may well outweigh it.

Finally, the approach of a common framework is to provide an accessible solution to some of the common problems faced by many scientific collaborations, as outlined in this report.
This generality is already being realized, with new production deployments in early stages, and more in development.
Long-term, we envision a community-driven project where any group deploys this system for their own use cases.
New features from one instance benefit others through the shared framework, while core development, centered around our needs as scientists and adapting to the latest technologies, is not duplicated across efforts and benefits all at once.


\bibliography{refs}

\appendix
\section{Automated Judge Prompt}
\label{sec:appendix_judge_prompt}

The four-judge panel and the source-free GLM-5.1 judge use the reference-free
prompt below, reproduced verbatim. Each judge sees the question and the
generated answer alone, with no reference answer; the \texttt{\{\{QUESTION\}\}}
and \texttt{\{\{GENERATED\_ANSWER\}\}} placeholders are filled per question. A
second variant, used only when a configuration returns retrieved sources, adds
a fifth source-faithfulness dimension on the same 1--5 scale; we report it
separately and exclude it from the four-dimension mean.

\begin{Verbatim}[breaklines=true,breakanywhere=true,fontsize=\footnotesize]
You are an expert evaluator for a CMS Computing Operations AI assistant. Evaluate the generated answer on the following dimensions using a 1-5 scale.

IMPORTANT PRINCIPLES:
- This is a REFERENCE-FREE evaluation. Score based on the answer's own quality alone.
- Unsupported specific claims (invented ticket numbers, dates, data values with no cited source) are WORSE than honest vagueness. An answer saying "I don't have access to look that up" is more trustworthy than one inventing data.
- Non-responses ("I'm ready to help!", empty answers, greetings) score 1 on all dimensions.
- ANTI-LENGTH BIAS: Do NOT reward longer answers for being longer. A concise, accurate answer should score as high or higher than a verbose answer that pads with generic advice. Score based on information quality, not quantity.

**Relevance** - Does the answer address the specific question that was asked?
- 5: Directly and precisely addresses the question - every part of the response is on-topic
- 4: Addresses the question with minor tangential content
- 3: Partially addresses the question but includes significant off-topic material, or only addresses part of a multi-part question
- 2: Mostly off-topic - touches on the general subject area but does not answer what was asked
- 1: Completely irrelevant, or a non-response ("I'm ready to help!", empty, greeting-only)

**Completeness** - How many aspects of the question does the answer address?
Assess scope by inferring what a full answer would need to cover from the question itself. For multi-part questions, a complete answer addresses all parts.
- 5: Addresses all aspects of the question - no significant gaps
- 4: Addresses most aspects, one minor gap
- 3: Addresses the core question but misses important context or sub-questions
- 2: Only partially addresses the question - significant gaps
- 1: Does not meaningfully address the question, or is a non-response

**Specificity** - Does the answer provide concrete, actionable details - or only vague generalities?
Concrete details include: specific commands, configuration values, ticket numbers, data values, step-by-step procedures, tool names with usage instructions, dates, error codes with explanations.

CRITICAL GUARDRAIL - unsupported specifics vs. honest vagueness:
An answer that provides specific details *grounded in cited sources or tool output* should score high. An answer that provides specific details *without any supporting evidence* (no citations, no tool output, no documentation references) should score LOWER than an answer that is honestly vague - because unsupported specifics may be fabricated and would mislead an operator.

- 5: Rich in concrete, well-supported details - commands, data, ticket references, step-by-step procedures grounded in sources or tool output
- 4: Provides useful specific details, mostly supported; minor unsupported claims
- 3: Mix of specific and vague - some actionable content but also generic advice ("check the logs", "contact the team")
- 2: Mostly vague or generic advice with little actionable content, OR provides unsupported specifics without any citations/evidence
- 1: Entirely vague ("look into it"), a refusal with no guidance, or a non-response

**Helpfulness** - Would a CMS computing operator be able to make progress on their task using this answer?
This is the bottom-line pragmatic dimension. An answer can be relevant, complete, and specific but still unhelpful if it points in the wrong direction.
- 5: An operator could act on this answer immediately - clear, correct next steps with enough detail to execute
- 4: Useful - provides a path forward, may require minor follow-up to fully act on
- 3: Somewhat useful - gives the operator a starting point but requires significant additional investigation
- 2: Minimally useful - vague pointers or a refusal with no alternative guidance
- 1: Not useful or actively harmful - would send the operator in the wrong direction, or is a non-response

Question:
{{QUESTION}}

Generated Answer:
{{GENERATED_ANSWER}}

Evaluate each dimension individually BEFORE assigning any scores. Think step-by-step about what the question asks, what the answer provides, and how well the answer serves an operator.

Return a JSON object with:
  - "reasoning": your step-by-step analysis (2-4 sentences)
  - integer scores (1-5) for each of: "relevance", "completeness", "specificity", "helpfulness"
\end{Verbatim}

\end{document}